\documentstyle[mnras_cite,psfig]{mn}

\def\gsim{~\rlap{$>$}{\lower 1.0ex\hbox{$\sim$}}}
\def\lsim{~\rlap{$<$}{\lower 1.0ex\hbox{$\sim$}}}

\def\HI{H{\sc i}}

\begin{document} 

\title{The Effects of Photoionization on Galaxy Formation --- II:
Satellite Galaxies in the Local Group}  
\author[A.~J.~Benson, C.~S.~Frenk, C.~G.~Lacey, C.~M.~Baugh \&
S.~Cole]{A.~J.~Benson$^1$, C.~S.~Frenk$^2$, C.~G.~Lacey$^3$,
C.~M.~Baugh$^2$ \& S.~Cole$^2$ \\ 
1. California Institute of Technology, MC 105-24, Pasadena, CA 91125,
U.S.A. (e-mail: abenson@astro.caltech.edu) \\ 
2. Physics Department, University of Durham, Durham, DH1 3LE, England \\
3. SISSA, Astrophysics Sector, via Beirut 2-4, 34014 Trieste, Italy}

\maketitle

\begin{abstract}
We use a self-consistent model of galaxy formation and the evolution
of the intergalactic medium to study the effects of the reionization
of the universe at high redshift on the properties of satellite
galaxies like those seen around the Milky Way. Photoionization
suppresses the formation of small galaxies, so that surviving
satellites are preferentially those that formed before the universe
reionized. As a result, the number of satellites expected today is
about an order of magnitude smaller than the number inferred by
identifying satellites with subhalos in high-resolution simulations of
the dark matter. The resulting satellite population has an abundance
and a distribution of circular velocities similar to those observed in
the Local Group. We explore many other properties of satellite
galaxies, including their gas content, metallicity and star formation
rate, and find generally good agreement with available data. Our model
predicts the existence of many as yet undetected satellites in the
Local Group. We quantify their observability in terms of their
apparent magnitude and surface brightness and also in terms of their
constituent stars. A near-complete census of the Milky Way's
satellites would require imaging to V$\approx 20$ and to a surface
brightness fainter than 26 V-band magnitudes per square
arcsecond. Satellites with integrated luminosity ${\rm V}=15$ should
contain of order 100 stars brighter than ${\rm B}=26$, with central
stellar densities of a few tens per square arcminute. Discovery of a
large population of faint satellites would provide a strong test of
current models of galaxy formation.\\

\end{abstract}

\section{Introduction}

High-resolution N-body simulations of the formation of dark matter
halos in the cold dark matter (CDM) cosmogony reveal a large number of
embedded subhalos that survive the collapse and virialization of the
parent structure \cite{klypin99a,moore99}.  Although their aggregate
mass typically represents less than 10$\%$ of the total halo mass,
these substructures are very numerous. In rich clusters, the abundance
of surviving subhalos is comparable to the abundance of bright
galaxies. In galaxy halos, on the other hand, the number of subhalos
exceeds the number of faint satellites observed in the Local Group by
well over an order of magnitude. This discrepancy has recently been
highlighted as a major flaw of the CDM cosmogony and has prompted
investigation of alternative cosmological models. These range from
non-standard models of inflation \cite{kamion99} to models in which
the universe is dominated by warm, self-interacting or annihilating
dark matter which may not generate small-scale substructure in
galactic halos, but may do so on cluster scales
\cite{hogan99,spergel99,moore00,yoshida00,craigdavis01}.

That hierarchical clustering theories predict many more small dark
matter halos than there are faint galaxies in the local universe has
been known for a long time, as has one possible solution to this
apparent conflict \cite{wr78}. Feedback generated by the energy
injected into galactic gas in the course of stellar evolution can
regulate star formation in small halos rendering their galaxies too
faint to be detected in local surveys
\cite{wr78,dekel86,cole91,wf91,lacey91}. The overabundance of dark
subhalos around the Milky Way predicted by CDM models was also known
before the high resolution N-body simulations highlighted the
discrepancy \cite{kwg93}. Using a semi-analytic model of galaxy
formation based on the extended Press-Schechter theory for the
assembly histories of dark halos \cite{bond91,bower91}, Kauffmann et
al. recognized the Milky Way ``satellite'' problem and examined
various possible solutions (see their Fig.~1). They concluded that the
best way to reconcile their models with the luminosity function of the
Milky Way's satellites was to assume that gas is unable to cool within
dark matter halos of circular velocity less than 150~km\,s$^{-1}$ at
redshifts between 5 and 1.5.  They suggested that this effect might
result from a photoionizing background at high redshift, but also
noted that the required suppression threshold of $150 {\rm km s^{-1}}$
was much larger than was expected based on physical calculations of
the effects of photoionization.  \scite{moore01} has expanded on the
reasons why standard supernova feedback of the kind invoked to explain
the relative paucity of faint field galaxies in CDM models does not,
on its own, solve the Milky Way satellite problem. He points out that
the observed satellites of the Milky Way of a given abundance have
circular velocities which are about three times smaller than the
circular velocities of subhalos of the same abundance in the N-body
simulations.

The lack of a Gunn-Peterson effect in the spectra of high redshift
quasars indicates that the Universe was reionized at $z\gsim 6$
\cite{fan00}. The reionization of the universe raises the entropy
of the gas that is required to fuel galaxy formation, preventing it
from accreting onto small dark matter halos and lengthening the
cooling time of that gas which is accreted. The inhibiting effects of
photoionization have been investigated in some detail
\cite{rees86,babul92,efstath92,shapiro94,katz96,quinn96,thoul96,abel97,kepner97,weinberg97,navarro97,barkana99},
with the conclusion that galaxy formation is strongly suppressed by
reionization in halos of circular velocity, $V_{\rm C}\lsim
60$~km\,s$^{-1}$. Although this value is smaller than the value
assumed by \scite{kwg93}, their general picture remains valid: the
number of satellites around bright galaxies is much smaller than the
number of dark matter subhalos because only those subhalos that were
already present before reionization were able to acquire gas and host
a visible galaxy.  This idea has recently been investigated further by
\scite{bullock00}. These authors followed the formation of dark halos
using a merger tree formalism similar to that of \scite{kwg93}, but
taking into account the tidal effects experienced by substructures
when they are accreted into their parent halo. Combining their halo
model with a simple argument based on the mass-to-light ratio of
satellite galaxies, they calculated their observability, concluding
that, for a reasonable redshift of reionization, the number of visible
satellites would indeed be close to that observed. A similar
conclusion was reached by \scite{somerville01} using a semi-analytic
model similar to that of \scite{cole2000} but with more limited model of
photoionization than used in this work (specifically, the model use by
\scite{somerville01} does not self-consistently evolve the properties
of galaxies and the IGM and does not account for the effects of
photoheating of virialized gas in halos or tidal disruption of
satellites).

In this paper, we investigate the abundance and properties of
satellite galaxy populations using a model of galaxy formation which
self-consistently calculates the physics of reionization and the
process of galaxy formation. We use the semi-analytic techniques
developed by \scite{cole2000} which we have recently extended to
enable calculation of the coupled evolution of the intergalactic
medium (IGM) and galaxies \cite{benson01bdum}. The model follows the
formation and evolution of stars, the production of ionizing photons
from stars and quasars, the reheating of the IGM, and the associated
suppression of galaxy formation in low mass halos. The model also
incorporates a detailed treatment of the dynamics of satellite halos
under the influence of dynamical friction and tidal forces. Like the
halo model of \scite{bullock00}, our model agrees with the results of
the high-resolution N-body simulations. In Paper~I, we demonstrated
that reionization reduces the number of faint field galaxies in the
local Universe, flattening the faint end slope of the galaxy
luminosity function, in good agreement with the most recent
observational determinations. In this paper, we will explore the
properties predicted by this very same model (i.e. with the same
parameter values) for the population of satellites around galaxies
like the Milky Way. We will carry out as detailed a comparison as is
possible with current observational data and present tests of our
model which rely on the prediction of a large and as yet undetected
population of faint satellites in the halo of the Milky Way.

The remainder of this paper is organised as follows. In
\S\ref{sec:model}, we briefly describe our model. In 
\S\ref{sec:abundance}, we calculate the expected luminosity function and
circular velocity function of satellites around galaxies like the 
Milky Way. In this section, we  also discuss observational strategies for
discovering the large faint satellite population predicted by our model. In
\S\ref{sec:otherprop}, we compare the gas content, star formation rate,
metallicity and structure of our model satellites with observational data.
Finally, in \S\ref{sec:discuss} we present our main conclusions.

\section{Model}
\label{sec:model}

We begin by briefly describing our model, a full description of which
may be found in Paper~I.  We represent the IGM as a distribution of
gas elements whose density evolves as a result of the expansion of the
Universe and the formation of structure. Each element `sees' a
background of ionizing photons emitted by stars (as given by the star
formation history in our model of galaxy formation) and by quasars (as
obtained from the observational parametrization of \pcite{madau99}),
which ionize and 
heat the gas. From this, we derive the thermal and ionization history
of the IGM. The hot IGM acquires a significant pressure which hampers
the accretion of gas into low mass dark matter halos. From the
inferred thermal state of the IGM we derive the filtering mass,
defined as the mass of a dark matter halo which accretes a gas mass
equivalent to 50\% of the universal baryon fraction \cite{gnedin2000b}, as
a function of time.

Knowledge of the filtering mass allows us to determine how much gas is
available for galaxy formation in a given dark matter halo at any
time. The increase in entropy produced by reionization causes lower
mass halos to contain a smaller fraction of their mass in the form of
gas than larger mass halos. The ionizing background which accumulates
after reionization also heats gas already present in dark matter
halos, preventing it from cooling into the star forming phase. We
include this heating in our estimates of the mass of gas which can
cool, resulting in a further suppression of galaxy formation. When a
galaxy falls into a larger halo becoming a satellite in it, we compute
its orbit through the halo in detail, accounting for the effects of
dynamical friction, tidal limitation and gravitational shocking (as
described by \scite{taylor00} and in Paper~I). As was demonstrated in
Paper~I, our model of satellite dynamics reproduces the results of
high-resolution N-body simulations with reasonable accuracy in both of
the cosmological models for which numerical results are available.

In Paper~I, we performed calculations using our extended galaxy
formation model in a $\Lambda$CDM cosmology (mean mass density
$\Omega_0=0.3$, cosmological constant term $\Lambda/3{\rm H}^2_0=0.7$,
mean baryon density $\Omega_{\rm b}=0.02$, and Hubble parameter
$h=0.7$.)  The values of the model parameters required to describe the
relevant physical processes are strongly constrained by a small subset
of the local galaxy data; we showed that the model also reproduces
many other properties of the galaxy population at $z=0$ (see also
\pcite{cole2000}). We will retain exactly the same parameter values
throughout the present work. The result is a fully-specified model of
galaxy formation that incorporates the effects of reionization and
tidal limitation and which we now use to explore in detail the
properties of satellite galaxies around the Milky Way.

\section{The abundance of satellites}
\label{sec:abundance}

We study the population of satellite galaxies that form in an ensemble of
halos harbouring galaxies similar to the Milky Way. We class a galaxy as
being ``similar to the Milky Way'' if the circular velocity of its disk 
$(V_{\rm C,d}$, measured at the disk half-mass radius) is between 210 and
230~km\,s$^{-1}$, and if its bulge-to-total mass ratio (including
stars and cold gas) is in the range 0.05 to 0.20 (which is 
approximately the range found by \scite{dehnen98} in mass models of the
Milky Way).

Using our model of galaxy formation, including all the effects of
photoionization and tidal limitation described in \S\ref{sec:model},
we construct 1800 realizations of dark matter halos with mass in the
range $4.0\times 10^{11}$ to $2.3\times 10^{12}h^{-1}M_\odot$ (the
range in which we find galaxies similar to the Milky Way) at $z=0$. We
ensure that our calculation resolves all halos into which gas is able
to accrete and cool in the redshift interval 0 to 25. From the set of
simulated halos we select those which contain a central galaxy similar
to the Milky Way.  We find approximately 70 such halos in our sample.
These have ``quieter'' merger histories than is typical for halos of
their mass, as was shown, for example, by
\scite{baugh96}. All other halos are discarded and, for the remainder
of this paper, we consider only the satellite populations of halos
hosting Milky Way type galaxies. We refer to this sample of satellites
as our ``standard model.'' For comparison, we also generated a sample
of Milky Way satellites using our model with no
photoionization (but still including the effects of tidal limitation of
satellites), which we will refer to as the ``no photoionization''
model, and a sample of Milky Way satellites with the original
model of \scite{cole2000}, which includes neither photoionization nor
tidal effects.

To test our model, we make use of observational data on Local Group
galaxies taken from the
compilation by \scite{mateo98}. Mateo points out that the census of
Local Group dwarfs is almost certainly incomplete. Unfortunately, the
observational selection effects are not quantitatively well
understood, and so we do not
attempt to correct for them here, but note that they could possibly
alter the observational results significantly. Some galaxies classed
as members of the Local Group lie outside the virial radii of the dark
matter halos thought to be associated with the Milky Way and M31. For
the Milky Way, we can estimate the virial radius using the
three-component mass models of \scite{dehnen98}, some of which assume
dark matter halos with the NFW \cite{nfw97} profile appropriate to the
CDM universe that we are considering here. Of these, their model 2d
gives the best fit to the observational data that they consider. 
Assuming a spherical top-hat collapse model for
the Milky Way halo then implies a virial mass and radius of $M_{\rm
MW}=1.11\times 10^{12}M_\odot$ and $R_{\rm MW}=$272~kpc respectively
(for the particular set of cosmological parameters considered in this
paper). This is in good agreement with the virial radii of the halos
that end up hosting Milky Way type galaxies in our model, which
typically have $R_{\rm vir}\approx 300$~kpc. (The model predicts a
distribution of $R_{\rm vir}$ because Milky Way type galaxies can be
found in halos with a range of masses). Lacking any better estimate,
we also take $R_{\rm MW}$ as the virial radius of M31, since the
measured circular velocity of M31 is quite close to that of the Milky Way. 
Throughout this
section, we will distinguish between Local Group satellites and
satellites lying within $R_{\rm MW}$ of either the Milky Way or M31.

\subsection{Abundance as a function of luminosity and stellar mass}
\label{sec:abundancelm}

\begin{figure}
\psfig{file=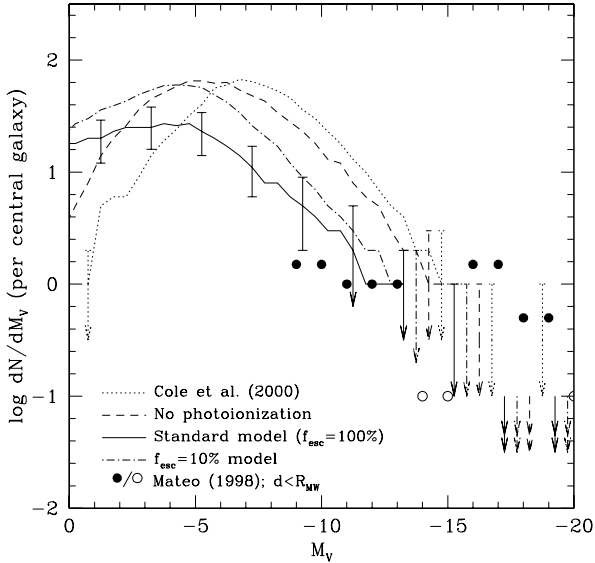,width=80mm}
\caption{The V-band luminosity function of satellite galaxies (per
central galaxy).  The Local Group data, taken from the compilation of
\protect\scite{mateo98}, are shown as filled circles, except for bins
in which the luminosity function is zero which we indicate by an open
circle at arbitrary position on the y-axis (faintwards of $M_{\rm
V}=-9$ there are no known satellites so we plot no symbols). Only
satellites within $R_{\rm MW}$ of the Milky Way or M31 are included
and no correction for any possible incompleteness has been applied.
The median luminosity function of satellites around pairs of Milky Way
type galaxies in our model is shown by the solid line, with errorbars
indicating the 10\% and 90\% intervals of the distribution of
luminosity functions found in approximately 70 realizations of the
satellite galaxy population. Where the 10\% interval corresponds to
zero galaxies, a single-headed downward-pointing arrow is
plotted. Where there is no connecting line, the median is zero, and
the 90\% interval is shown by an errorbar and arrow.  Where the 90\%
interval corresponds to zero galaxies (i.e. less than one in ten of
the simulated halos contained any galaxies of this magnitude), we show
a double-headed downward-pointing arrow (at an arbitrary position on
the y-axis). The dashed line corresponds to the model in which the
effects of photoionization are ignored (but tidal limitation of
satellites is included), while the dotted line shows the prediction
from the model of \protect\scite{cole2000} and the dot-dashed line
shows results from our standard model with the reduced escape fraction
of $f_{\rm esc}=10\%$. For clarity we have omitted error bars from
these three models; these are typically 10--20\% smaller than for the
standard model.}
\label{fig:satLF}
\end{figure}

We begin by comparing the luminosity function of the population of
satellites in the model with the observed luminosity function of
satellites of the Local Group. Fig.~\ref{fig:satLF} shows the V-band
luminosity function of satellites\footnote{We include the SMC, LMC and
M33 in this category.}, normalised to the number of central galaxies
(i.e. the Milky Way and M31), as filled circles. We include only
satellites within $R_{\rm MW}$ of either the Milky Way or M31. 
Throughout this paper, we use magnitudes corrected for foreground
extinction by the Milky Way using the reddening values listed by
\scite{mateo98}. The median luminosity function from all pairs of
Milky Way type galaxies in our standard model is shown by the solid
line, with errorbars enclosing 10\% and 90\% of the distribution of
luminosity functions. (We calculate the median and intervals for pairs
of halos since we are comparing to the combined Milky Way and M31
datasets and the scatter in model predictions for pairs of halos is,
of course, smaller than for single halos.) The satellite galaxies in
our model typically have only very small amounts of internal
dust-extinction (all but a tiny fraction have less than 0.1 magnitudes
of extinction in the V-band). Nevertheless, we include the effects of
internal dust-extinction in all model magnitudes.

It is immediately apparent in Fig.~\ref{fig:satLF} that
photoionization does help to reduce the number of satellite galaxies:
our model predicts many fewer satellites than the model of
\scite{cole2000}. Furthermore, the dashed line shows that tidal stripping
also acts to reduce the number of satellites of a given luminosity,
but this is a much smaller effect. For $M_{\rm V}\approx -10$, the
combined effects of tidal stripping and photoionization reduce the
number of satellites by about a factor of 10 compared to the
\scite{cole2000} model. Note that the model predicts significant
variation in the satellite luminosity function from halo to halo. Once
this scatter is taken into account, our standard model is in
reasonable agreement with the Local Group luminosity function, except
perhaps at the brightest magnitudes, where the model underpredicts the
number of satellites. However, the statistics are 
poor at these magnitudes, because there are so few galaxies. For
example, the $-17.5 \geq M_{\rm V} > -18.5$ bin in 
Fig.~\ref{fig:satLF} contains a single galaxy (the LMC) within $R_{\rm
MW}$, implying a mean of $0.5$ such satellites per central galaxy. Our
model predicts a mean of approximately 0.08. Since no halo in the
sample contains more than one such galaxy, 8\% of the model halos
contain a satellite galaxy in this luminosity range. Thus, such bright
satellites are not impossible in our model, but they are quite
rare. 

As was noted in Paper~I, the disk velocity dispersion used in the
\scite{taylor00} model ($\sigma_{\rm d}=V_{\rm C,d}/\sqrt{2}$) is
rather high for the Milky Way. This quantity determines the magnitude
of the dynamical friction force felt by satellites which pass close to
the disk of the central galaxy, and so may affect the rate of merging
for satellites on highly eccentric orbits. An overly large
$\sigma_{\rm d}$ would result in a reduced dynamical friction force
and a lower rate of mergers, potentially leading us to overestimate
the number of satellites remaining. However, adopting the more
appropriate value, $\sigma_{\rm d}=0.2 V_{\rm C,d}$, makes virtually
no difference to the predicted number of satellites, since the the
disk velocity dispersion is only important for satellites very close
to the central galaxy, and which therefore tend to merge shortly
afterwards in any case. On the other hand, the abundance of the
faintest satellites is fairly sensitive to two uncertain model
assumptions, the escape fraction of ionizing photons, $f_{\rm esc}$,
and the luminosity of the galaxies that we have classed as resembling
the Milky Way.

In Paper~I, we found that adopting an escape fraction, $f_{\rm
esc}=100\%$ (i.e. all ionizing photons produced by stars escape from
their galaxy into the IGM), results in a reasonable redshift of
reionization ($z \approx 8$), but also in an ionizing background which
is significantly higher than current observational estimates. On the
other hand, a fraction, $f_{\rm esc}=10\%$, results in a more
reasonable ionizing background and is in better agreement with
observational determinations of $f_{\rm esc}$
\cite{leitherer95,steidel00}, but implies a lower redshift of
reionization ($z \approx 5.5$), perhaps uncomfortably close to the
lower limit imposed by the lack of a Gunn-Peterson effect in the
highest redshift quasars. We find that the lower redshift of
reionization associated with $f_{\rm esc}=10\%$ leads to a larger
number of satellites of a given luminosity (as shown by the dot-dashed
line in Fig.~\ref{fig:satLF}), producing a luminosity function which,
faintwards of $M_{\rm V}=-11$, is a factor of about two too high
compared with the data (even after allowing for halo-to-halo
variations).  However, possible incompleteness at the faint end of the
observed luminosity function makes this comparison inconclusive.  We
will adopt $f_{\rm esc}=100\%$ as our standard value for all results
presented in this paper, unless stated otherwise.

A second source of uncertainty is the identification of model galaxies
with the Milky Way.  As discussed in Paper~1 (see Fig.~9), our
modelling of the structure of disks is crude and produces circular
velocities for spiral galaxies that are typically 30\% larger for
their luminosity than is implied by the observed Tully-Fisher
relation. As a result, the model galaxies that we have identified with
the Milky Way according to their circular velocity are somewhat
fainter than galaxies lying on the mean Tully-Fisher relation at that
circular velocity. We could alternatively select Milky Way galaxies
according to luminosity. Direct estimates of the luminosity of the
Milky Way are rather uncertain and so we instead estimate the
luminosity from the observed Tully-Fisher relation as advocated by
Binney \& Merrifield (1998, \S10.1). The observed Tully-Fisher
relation \cite{matthew92} implies an I-band magnitude in the region of
$-21.5 \gsim M_{\rm I}-5\log h \gsim -22.0$ for a typical galaxy with
the circular speed of the Milky Way. Model galaxies with I-band
magnitude in this range (and bulge-to-total ratio in the range
discussed above) typically have higher circular velocities, and
inhabit higher mass halos than the Milky Way type galaxies of our
original sample. As a result, this new sample has a substantially
larger number of satellites. Its luminosity function is compatible
with observations for magnitudes brighter than $M_{\rm V}\approx -13$,
but faintwards of this, it overpredicts the number of satellites.
Again, possible incompleteness in the faint data makes it difficult to
exclude this model, but for it to be compatible with observations, the
data would have to be severely incomplete so that at least 70\% of
$M_{\rm V}\approx -10$ satellites should have been missed.

\begin{figure}
\psfig{file=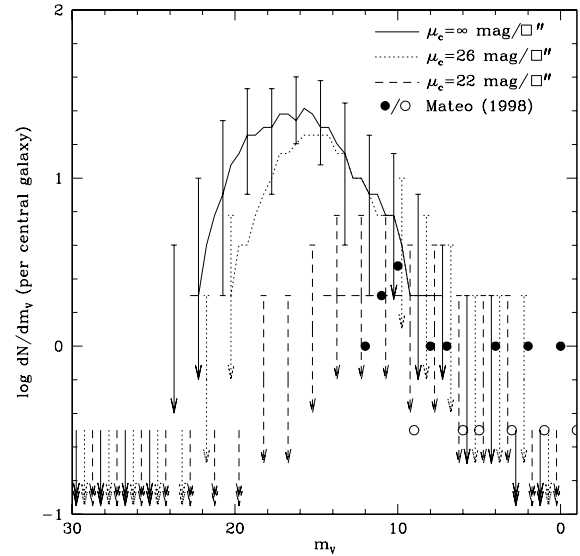,width=80mm}
\caption{V-band number counts of satellites.  The solid line shows the
results from our standard model with no cut on central surface
brightness, while dotted and dashed lines show the effects of applying
a central surface brightness cut of 26 and 22 magnitudes per square
arcsecond respectively. Lines show the median counts in the standard
model, with errorbars indicating the 10\% and 90\% intervals of the
distribution for the results with no surface brightness cut. (The
errorbars for the 26 magnitudes per square arcsecond sample are
comparable to these, except for $m_{\rm V}<16$ where they are about
twice as large.) Where the 10\% interval corresponds to zero galaxies,
a single-headed downward pointing arrow is shown. Where only an
errorbar or arrow is shown (i.e. there is no connecting line), the
median is zero, and where the 90\% interval corresponds to zero
galaxies we show a double-headed downward-pointing arrow (at arbitrary
position on the y-axis). Note that unlike the luminosity function
(Fig.~\protect\ref{fig:satLF}), the results here refer to a single
halo (as opposed to pairs of halos). Apparent magnitudes for
satellites are computed from their position in the host halo, assuming
the observer is located 8~kpc from the centre. The filled circles show
the observed counts from \protect\scite{mateo98}, including only those
galaxies within $R_{\rm MW}$ of the Milky Way. Where a bin contains no
observed satellites we show an open circle at arbitrary position on
the y-axis (faintwards of the $m_{\rm V}=12$ bin there are no observed
satellites, so we do not plot any symbols).}
\label{fig:satapp}
\end{figure}

As can be clearly seen from Fig.~\ref{fig:satLF}, our model predicts
that the satellite galaxy luminosity function should continue to rise
at magnitudes fainter than the faintest satellite yet observed. The
luminosity function peaks at $M_{\rm V}\approx -3$ and then falls off
at fainter magnitudes. This cut-off is caused by the inability of gas
below $10^4$K to cool when no molecular hydrogen is present. The cut
off occurs at significantly fainter magnitudes in our standard model
than in the model with no photoionization or the \scite{cole2000}
model since photoionization reduces the efficiency of galaxy formation
in halos of a given mass (or, equivalently, virial temperature) making
the galaxies in those halos fainter\footnote{We are able to predict
the properties of these very faint galaxies by extrapolating our
standard rules for star formation, feedback etc. to these very small
objects. However, it should be kept in mind that these very faint
galaxies typically contain only a few thousand Solar masses of
stars. It is not clear how well our simplified rules for star
formation describe reality in such systems, where the entire galaxy is
less massive than a single giant molecular cloud.}. It is interesting
therefore to consider the observability of the galaxies that make up
the faint end of the satellite luminosity function. In
Fig.~\ref{fig:satapp} we show the number of satellite galaxies per
halo as a function of their apparent V-band magnitude. To infer the
distances to the satellites, we have made use of their orbital
positions, assuming that they are observed from a location 8~kpc from
the centre of the host halo (i.e. at about the position of the Sun in
the Milky Way). (The mean distance for $M_{\rm V}\leq-10$ satellites
is 95~kpc and this changes only slowly with $M_{\rm V}$)\footnote{The
number density of satellites within the virial radius of a Milky Way
type halo in our model scales roughly as $r^{-2}$ for $r\gsim 10$~kpc
with a rapid cut-off towards $R_{\rm vir}$, but flattens significantly
at smaller radii since these inner satellites are strongly affected by
tidal forces and dynamical friction.}.  The solid line shows all
satellites, while the dotted and dashed lines show only those whose
central surface brightness is brighter than 26 and 22 magnitudes per
square arcsecond respectively. Central surface brightnesses for model
galaxies were computed using the predicted luminosities and sizes,
with disk components modelled as exponential disks, and spheroids as
King profiles with $r_t/r_c=10$. The faintest known Local Group
galaxy is Tucana ($m_{\rm V}=15.15$ with almost no foreground
extinction; note that this point is not seen in Fig.~\ref{fig:satapp}
since Tucana lies outside of the Milky Way's virial radius), only
slightly brighter than the peak in the predicted apparent magnitude
distribution. Surface brightness adds a further limit to the
detectablility of satellites. Currently the lowest surface brightness
member of the Local Group that is known is Sextans, with
$\mu_0=26.2\pm0.5$mag arcsec$^{-2}$ \cite{mateo98}. This is close to
the dotted line in Fig.~\ref{fig:satapp}. It is clear from
Fig.~\ref{fig:satapp} that the faint satellites have very low surface
brightness and that finding them will require very deep photometry.

\begin{figure}
\psfig{file=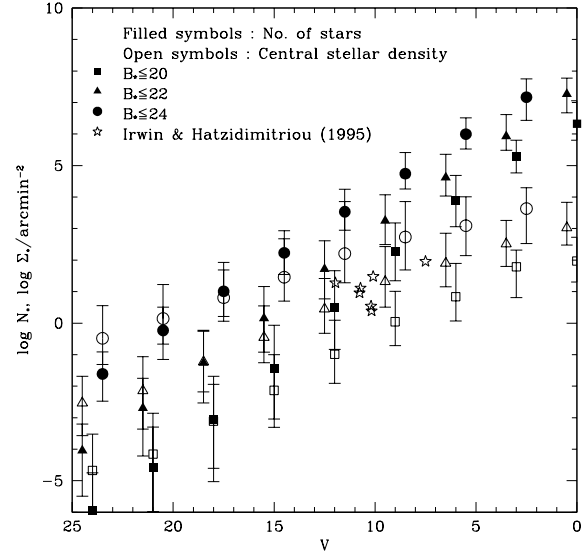,width=80mm}
\caption{The number of stars (filled symbols) and the surface density
of stars (open symbols) brighter than B$_\star$=20, 22 and 24
(squares, triangles and circles respectively) for satellites of Milky
Way type galaxies as a function of the apparent magnitude of the
galaxy. The measured central stellar densities for a subset of the
Local Group satellites measured by Irwin \& Hatzidimitriou (1995),
with a limiting magnitude $B=22$, are shown as open stars.}
\label{fig:satnstars}
\end{figure}

Several satellites have now been detected as an excess of stars
against the background of the Milky Way (e.g. \pcite{irwin95}). From
the star formation histories of our model satellites, we can calculate
the number and surface density of stars visible today. We use the
stellar evolutionary tracks of \scite{lejeune01} to obtain the number of
stars brighter than a certain luminosity, $L_{\rm c}$, in a single
stellar population of unit mass, as a function of age and metallicity:
\begin{equation}
N_{L_{\rm c}}(t,Z)=\int_{M_{\rm min}}^{M_{\rm max}} S_{L_{\rm
c}}(M,t,Z) {{\rm d}n \over {\rm d}M} {\rm d}M, 
\end{equation}
where ${\rm d}n/{\rm d}M$ is the stellar Initial Mass Function (IMF;
we use the IMF of \scite{kennicutt83}, as assumed in our calculations
of galaxy luminosities), normalized to unit mass and with minimum and
maximum masses, $M_{\rm min}$ and $M_{\rm max}$ (0.1 and 125$M_\odot$
respectively), and
\begin{equation}
S_{L_{\rm c}}(M,t,Z) = \left\{ \begin{array}{ll} 0 & \mbox{if }
L(M,t,Z)<L_{\rm c} \\ 1 & \mbox{if } L(M,t,Z)\geq L_{\rm c},
\end{array} \right. 
\end{equation}
where $L(M,t,Z)$ is the luminosity of a star of zero-age mass $M$, age
$t$ and metallicity $Z$. The number of stars more luminous than
$L_{\rm c}$ in a galaxy at the present day is then given by
\begin{equation}
N=\int_0^{t_0} \dot{\rho}_\star (t) N_{L_{\rm c}}(t_0-t,Z[t]) {\rm d}t,
\end{equation}
where $\dot{\rho}_\star(t)$ is the star formation rate in the galaxy
at time $t$ and $Z[t]$ is the metallicity of the stars being formed at
time $t$. For each satellite, we compute the luminosity corresponding
to a given apparent magnitude limit at its position in the halo and
determine $N$. From the known size of the disk and spheroidal
component of each model galaxy, we also derive the central number
density of these stars. Fig.~\ref{fig:satnstars} shows the result of
this calculation. (We do not include the effects of dust extinction on
the stellar luminosities since this varies across the galaxy, but we
do include them in the total galaxy magnitude. As noted above, these
internal extinction corrections are typically small in any case.) 
The central stellar densities
predicted by our model are in excellent agreement with those measured
by \scite{irwin95} from scanned photographic plates with a limiting
magnitude of B$=22$: compare the open stars with the open triangles in
Fig.~\ref{fig:satnstars}. Imaging two magnitudes deeper (circles)
would allow the detection of galaxies to V=17--18 at the same central
surface density, although the higher density of background objects may
make detection more difficult.

\subsection{Abundance as a function of circular velocity} 
\label{sec:abundancevc}

In this subsection we derive the abundance of model satellites as a
function of their circular velocity and compare it to the Local Group
data. This is the statistic originally employed by \scite{klypin99a}
and \scite{moore99} to highlight the apparent discrepancy between the
CDM cosmogony and the properties of the satellites of the Milky
Way. Their motivation for using the distribution of circular
velocities of satellite halos it is that can be straightforwardly
obtained from N-body simulations including only dark matter,
whereas the luminosities of satellite galaxies, while fairly easy to
measure observationally, cannot be so easily predicted theoretically,
because of the complicated dependence on the processes of cooling,
star formation, feedback and photoionization. The main drawback of
making the comparison in terms of circular velocities is that
measuring circular velocities is a difficult task for the majority of
the Local Group satellites which do not have gas disks, and in any
case these measurements give the circular velocity within the visible
galaxy, not the value at the peak of the rotation curve of the dark
halo, which is typically what is measured in the simulations. In
addition, the comparisons by \scite{klypin99a} and \scite{moore99}
implicitly assumed a tight relation between subhalo circular
velocity and satellite luminosity, since otherwise the observational
selection on luminosity modifies the form of the circular velocity
distribution, as we will see below.

Before proceeding with the calculation of the satellite circular
velocity distribution, it is important to check that our model
produces realistic sizes for the satellites, as well as the correct
luminosity-circular velocity relation. The size of a galaxy influences
the circular velocity because it determines the self-gravity of the
visible component (and the associated compression of the halo which we
treat using adiabatic invariants, as discussed in
\pcite{cole2000}). It also determines the part of the rotation
curve which is accessible to observations.  The luminosity-circular
velocity relation enters because, in practice, the distribution of circular
velocity is measured for a sample selected to be brighter than a given
luminosity.

\begin{figure}
\psfig{file=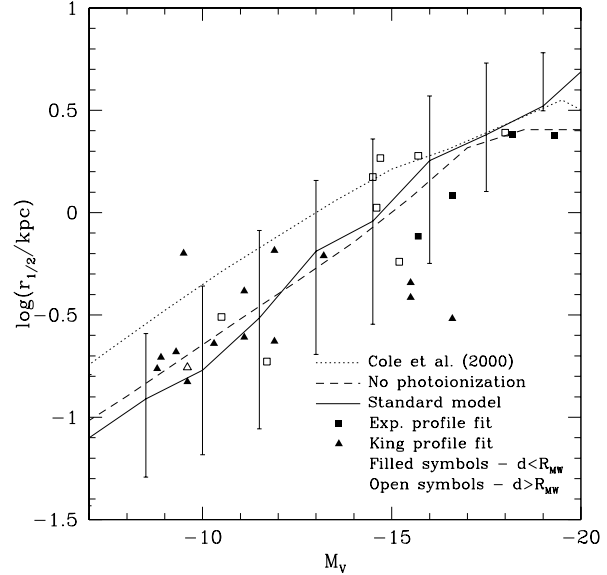,width=80mm}
\caption{The half-light radii of satellite galaxies as a function of
absolute V-band magnitude. The points show the radii of the satellites
of the Local Group, estimated from the data compiled by
\protect\scite{mateo98}. For spiral and irregular galaxies, we convert
their measured exponential scalelength, $r_{\rm exp}$, to a half-light
radius by assuming an exponential disk density profile,
$r_{1/2}\approx 1.68 r_{\rm exp}$; we show these points as
squares. For ellipticals and dwarf spheroidals, we use the quoted
parameters of the \protect\scite{king66} model fits to reconstruct the
density profile of the galaxy, and so infer the half-light radius. The
core and tidal radii listed by \protect\scite{mateo98} are measured
along the major axis of each galaxy. We replace these with the
equivalent radii for spherically symmetric systems by computing the
geometric mean of the radii along the semi-major and semi-minor axes
(using the measured ellipticities, which we assume to be independent
of radius).  These points are shown as triangles.  (In the few cases
where a galaxy is classed as Irr/dSph, we include it in the Irr class
for the purposes of this plot.) Filled points indicate satellites
within a distance $R_{\rm MW}\approx 270$~kpc from the Milky Way or
M31, and open symbols indicate more distant satellites.  The solid
line corresponds to our standard model, the dashed line to the model
that ignores photoionization but which includes tidal stripping of
satellite halos, and the dotted line to the \protect\scite{cole2000}
model.  The lines show the median relation and the errorbars indicate
the $10\%$ and $90\%$ intervals of the distribution in the standard
model. In this and subsequent figures errorbars are only shown where
the magnitude bin contains enough model galaxies to estimate the 10\%
and 90\% intervals accurately, and so do not appear at the brightest
magnitudes. The scatter in the no-photoionization model is similar to
this, but in the \protect\scite{cole2000} model it is typically
60-70$\%$ smaller. The half-light radius of the model satellites is
taken to be the half-mass radius of the galaxy (including both disk
and spheroid) remaining within the effective tidal radius. The
half-mass radii of the disk and spheroid prior to tidal limitation
were determined in the manner described by \protect\scite{cole2000}.}
\label{fig:sat_sizes}
\end{figure}

In Fig.~\ref{fig:sat_sizes} we compare the half-light radii of our
model satellites with observations. We estimate the half-light radii
of real satellites using published fits to their surface brightness
profiles (typically exponentials or King models) in the manner
described in detail in the figure caption\footnote{For reference, half
of the Local Group satellites listed by \scite{mateo98}, as well as
the LMC, SMC and M33, are irregulars and half are spheroidals; for
galaxies brighter than $M_{\rm V}=-8$, the model predicts 78\%
spiral/irregulars and 22\% spheroidals. It is likely that the
morphological evolution of satellites has been influenced by dynamical
processes not included in our model (e.g. \pcite{mayer01}).}.  For the
model satellites, we take the half-light radius to be the half-mass
radius of the stellar system (including both disk and spheroid) that
remains within the effective tidal radius. (In Paper~I we defined the
effective tidal radius as the radius in the original satellite density
profile beyond which material has been lost.) We show results for our
standard, no photoionization and \scite{cole2000} models. The sizes of
satellites in our standard model are significantly smaller than in the
\scite{cole2000} model, particularly fainter than $M_{\rm V}=-15$. As
the figure shows, much of this reduction is due to tidal limitation
which strips away the outer layers of the galaxies.  Photoionization
appears to have little effect on the sizes of satellites. In reality,
photoionization does reduce the size, but it also reduces the
luminosity, essentially preserving the size-luminosity relation, so
that the galaxies typically move in a direction almost parallel to the
median relation in Fig.~\ref{fig:sat_sizes}.  The model predictions
are in good agreement with the data and, given the rather crude
observational determinations available, this level of agreement seems
sufficient at present.

\begin{figure}
\psfig{file=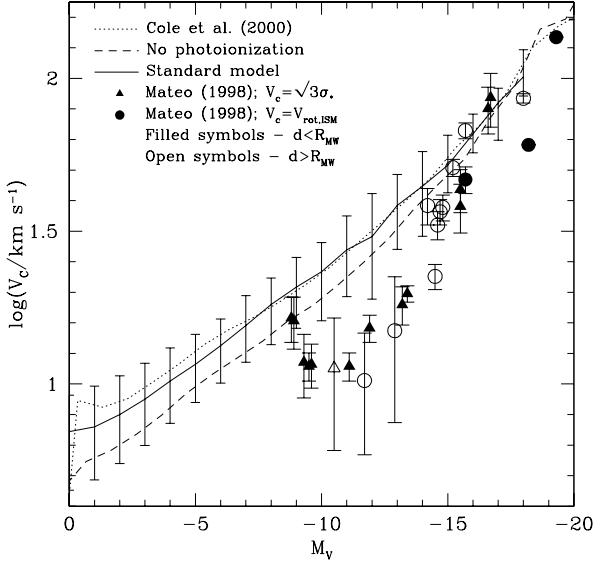,width=80mm}
\caption{The relation between circular velocity and V-band absolute
magnitude for satellite galaxies. For the real satellites of the Local
Group \protect\cite{mateo98}, circles indicate circular velocities
inferred from the rotation speed of the ISM, while triangles indicate
circular velocities estimated from the stellar velocity dispersion as
discussed in the text. Filled symbols denote galaxies within $R_{\rm
MW}$ of the Milky Way or M31, while open symbols denote more distant
galaxies. The median relation from our standard model is shown by the
solid line with errorbars indicating the 10\% and 90\% intervals of
the distribution. For the model, we plot the circular velocity at the
half-mass radius of the satellite (including both disk and spheroid
and taking only the mass remaining within the effective tidal
radius). The dashed line shows the results from the model in which the
effects of photoionization are ignored, but tidal limitation of halos
is included, while the dotted line shows the results from the model of
\protect\scite{cole2000}. The scatter in the former is similar to that
in the standard model, but in the latter it is about 70\% of that in
the standard model.}
\label{fig:satTF}
\end{figure}

The `Tully-Fisher' relation between the circular velocity and the
absolute V-band magnitude of satellites is plotted in
Fig.~\ref{fig:satTF}. Whenever possible, the circular velocity of real
satellites was estimated from the rotation velocity of gas in the ISM
(corrected for inclination). Where a measurement of this is not
available, we used instead the measured stellar velocity dispersion
multiplied by a factor $\sqrt{3}$.\footnote{The factor $\sqrt{3}$
follows from the assumption of isotropic stellar orbits in an singular
isothermal potential, for a stellar distribution with an $r^{-3}$
density profile, assumptions which may not be relevant to real
galaxies. Alternatively, the average line-of-sight velocity dispersion
of the satellite's spheroid can be estimated by modelling the spheroid
as a King profile (which is often a good fit to observed data) and
then solving the Jeans equation (see Paper~I, eqn.~21).  Assuming
isotropic orbits, we find $V_{\rm bulge}\approx 1.2 \sigma_*$ for
model satellites, albeit with large scatter, where $V_{\rm bulge}$ is
the circular velocity of the bulge at the stellar half-mass radius and
$\sigma_*$ is the stellar line-of-sight velocity dispersion. Adopting
this latter value does not change our conclusions for the velocity
function.}  For the model satellites, we plot the circular velocity at
the stellar half-mass radius, which is typically $\sim25$\% smaller
than the peak circular velocity in the tidally limited halo.  For the
fainter satellites ($M_{\rm V}\simeq -10$), the circular velocity at
the half-mass radius is about half the value at the virial radius,
whereas for the brighter satellites ($M_{\rm V}\simeq -15$), the two
are similar. The theoretical relations provide a good description of
the data, except in the magnitude range $-9 \lsim M_{\rm V} \lsim -14$
for which the model velocities are about a factor of 2 too high. In
this range, the no photoionization model performs slightly better than
the standard model since tidal limitation results in the rotation
velocity being measured at a smaller radius, an effect which is offset
by the fact that photoionization requires galaxies of fixed luminosity
to form in more massive halos when photoionization is switched on
(compare the solid and dashed lines in Fig.~\ref{fig:satTF}).  While
it is possible that this discrepancy may reflect a shortcoming of the
model, the estimation of circular velocities from current
observational data is highly uncertain; improved determinations are
extremely important. In addition, there is also some theoretical
uncertainty in the calculation of the circular velocity. For example,
as noted in Paper~I, our model does not account for changes in the
density profile of the satellite after tidal limitation, which may
reduce the rotation speed somewhat \cite{mayer01}.

\begin{figure}
\psfig{file=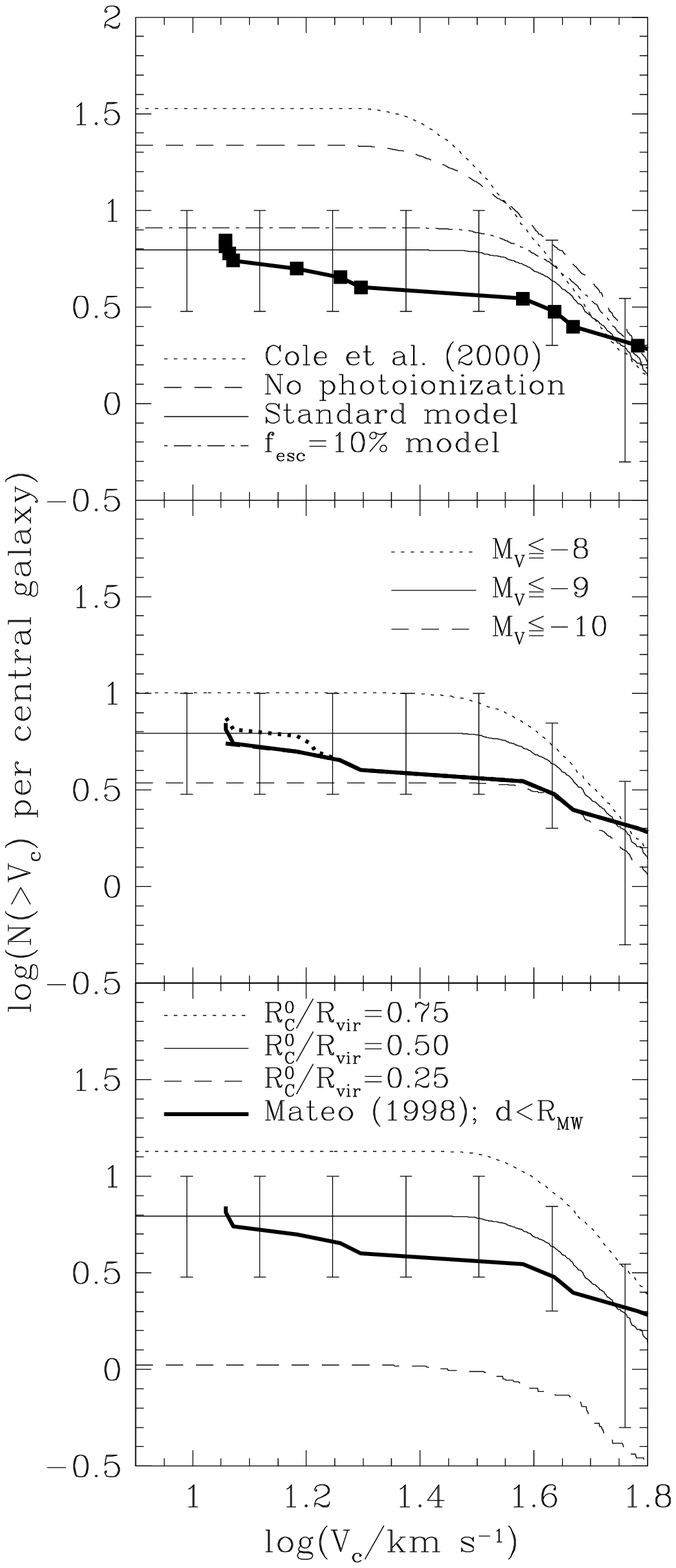,width=80mm,bbllx=0mm,bblly=15mm,bburx=110mm,bbury=265mm,clip=}
\caption{}
\end{figure}

\begin{figure}
\addtocounter{figure}{-1}
\caption{\emph{(cont.)} The cumulative velocity function of satellite
galaxies. Data for the Local Group per central galaxy (i.e. satellites
within $R_{\rm MW}$ of the Milky Way or M31), taken from
\protect\scite{mateo98}, are shown by heavy lines without errorbars
(in the top panel we also plot squares to indicate the contribution
from each individual satellite). \emph{Top panel:} The mean cumulative
velocity function of satellites brighter than $M_{\rm V}=-9$ in our
standard model is shown by the solid line, with errorbars enclosing
10\% and 90\% of the distribution of velocity functions (from a sample
of approximately 70 Milky Way type halos).  The circular velocities
plotted are measured at the half-mass radius of the combined disk and
spheroid system after accounting for tidal limitation. The dot-dashed
line shows the effect of reducing $f_{\rm esc}$ to 10\% (from 100\% in
the standard model). The dashed line shows the model in which the
effects of photoionization are neglected but satellite limitation by
tidal stripping is included, while the dotted line corresponds to the
model of \protect\scite{cole2000}. The scatter in the former is
similar to, and in the latter it is typically 70\% of, the scatter in
the standard model. All models have $R_{\rm c}^0/R_{\rm
vir}=0.5$. \emph{Middle panel:} Dependence on the absolute magnitude
cut. The thin lines show results for the standard model for the
different magnitude cuts: solid for $M_{\rm V}=-9$ (as in the top
panels), dotted for $M_{\rm V}=-8$, and dashed for $M_{\rm V}=-10$.
The thick lines show the Local Group data for the same magnitude cuts,
with the same line types (the $M_{\rm V}=-10$ line lies almost
entirely under the $M_{\rm V}=-9$ line).  \emph{Lower panel:} As
middle panel for the standard model but with different values of
$R_{\rm c}^0/R_{\rm vir}$ (as given in the legend); $R_{\rm
c}^0/R_{\rm vir}$ is the model parameter that controls the initial
orbital radii of satellites as defined in Paper~I.}
\label{fig:satMW}
\end{figure}

The satellite cumulative velocity function is shown in
Fig.~\ref{fig:satMW}. In our model of photoionization, every dark
matter halo whose virial temperature is larger than the minimum
temperature for cooling (see Fig.~1 of Paper~I) forms a galaxy
(although in low mass halos the galaxies have extremely low mass), and
tidal stripping never completely destroys a satellite halo (although
halos may be stripped to very small radii). As a consequence, if we
naively constructed the full cumulative velocity function of
satellites in our model, it would be identical to that of
\scite{cole2000}. In reality, many of the satellites in our model are
extremely faint either because photoionization severely restricted
their supply of star-forming gas, or because tidal stripping removed
most of their stars, and so would be unobservable. We therefore
apply an absolute magnitude limit to our sample of satellites and
construct the cumulative velocity function only for galaxies brighter
than this limit. The sample of known Local Group satellites does not have a
well-defined absolute magnitude limit (nor an apparent magnitude limit
for that matter), but the faintest galaxy listed by \scite{mateo98},
Draco, has $M_{\rm V}=-8.8$. We therefore choose $M_{\rm V}=-9$ as a
suitable cut for our models and apply the same cut to the data.

Results for our standard, no photoionization and \scite{cole2000}
models are shown with solid, dashed and dotted lines respectively in
Fig.~\ref{fig:satMW}.  It is immediately apparent that while the model
of \scite{cole2000} substantially overpredicts the number of
satellites, our photoionization model is in very good agreement with
the data. As the figure clearly shows, the main reduction comes from
the effects of photoionization; tidal limitation alone has only a
minor effect on the velocity function. The scatter from realization to
realization is large: the central 80\% of the distribution of velocity
functions spans a factor of 3. Adopting a magnitude cut at $M_{\rm
V}=-10$ produces equally good agreement (except, perhaps, at the
lowest circular velocities, $\lsim 13$km/s), but taking $M_{\rm V}=-8$
leads to roughly twice as many satellites in the range $16\hbox{km/s}
\lsim V_c \lsim 40\hbox{km/s}$ as is observed. In all cases, the
photoionization model predicts far fewer satellites than the
\scite{cole2000} model. Our results are weakly dependent on the
assumed photon escape fraction. The result of assuming $f_{\rm
esc}=10\%$ (instead of $f_{\rm esc}=100\%$ in the standard model) is
shown by the dot-dashed line in the upper panel. The number of
satellites with $V_c \lsim 30 {\rm km s^{-1}}$ is increased by
about 25\% in this latter case.

Finally we show in the lower panel of Fig.~\ref{fig:satMW} the effect
of varying the ratio $R_{\rm c}^0/R_{\rm vir}$, which parameterizes
the initial orbital energy of a satellite in terms of the energy of a
circular orbit of radius $R_{\rm c}^0$. In Paper~I, we demonstrated
that a value of $R_{\rm c}^0/R_{\rm vir}=0.5$ provides a good match to
the number of subhalos seen in high-resolution N-body simulations, and
is comparable to the mean value measured for satellites in those
simulations at $z=0$.  Increasing $R_{\rm c}^0/R_{\rm vir}$ to 0.75
produces somewhat too many satellites, while reducing it to 0.25
seriously underpredicts the number of satellites. Increasing $R_{\rm
c}^0/R_{\rm vir}$ makes satellite orbits begin at larger radii where
tidal forces are weaker and so allows more satellites to survive
(decreasing $R_{\rm c}^0/R_{\rm vir}$ has the opposite effect). The
magnitude of the changes induced by adopting these alternative values
for $R_{\rm c}^0/R_{\rm vir}$ is comparable to that seen for
calculations of satellite halo abundances in the pure dark matter
calculations reported in Paper~I, and so can be seen to be due almost
entirely to the enhanced(reduced) tidal limitation and dynamical
friction resulting from a smaller(larger) value of $R_{\rm c}^0/R_{\rm
vir}$, supplemented slightly by changes in the luminosities of
satellites which alter whether they meet our selection criteria for
this figure.

The exact distribution of $R_{\rm c}^0/R_{\rm vir}$ is therefore very
important, and would be worth determining accurately from numerical
simulations. We stress that the value $R_{\rm c}^0/R_{\rm vir}=0.5$
was chosen to match the results of N-body simulations of dark matter,
and also results in good agreement with the abundances of satellite
galaxies.

\section{Other properties of satellites}
\label{sec:otherprop}

In addition to the abundance, our model of galaxy formation predicts
many other properties of the satellite galaxy population and their
associated dark halos. We now explore some of them, particularly those
for which observational data are available.

\subsection{Gas content, star formation rates, metallicities and colours}

\begin{figure}
\psfig{file=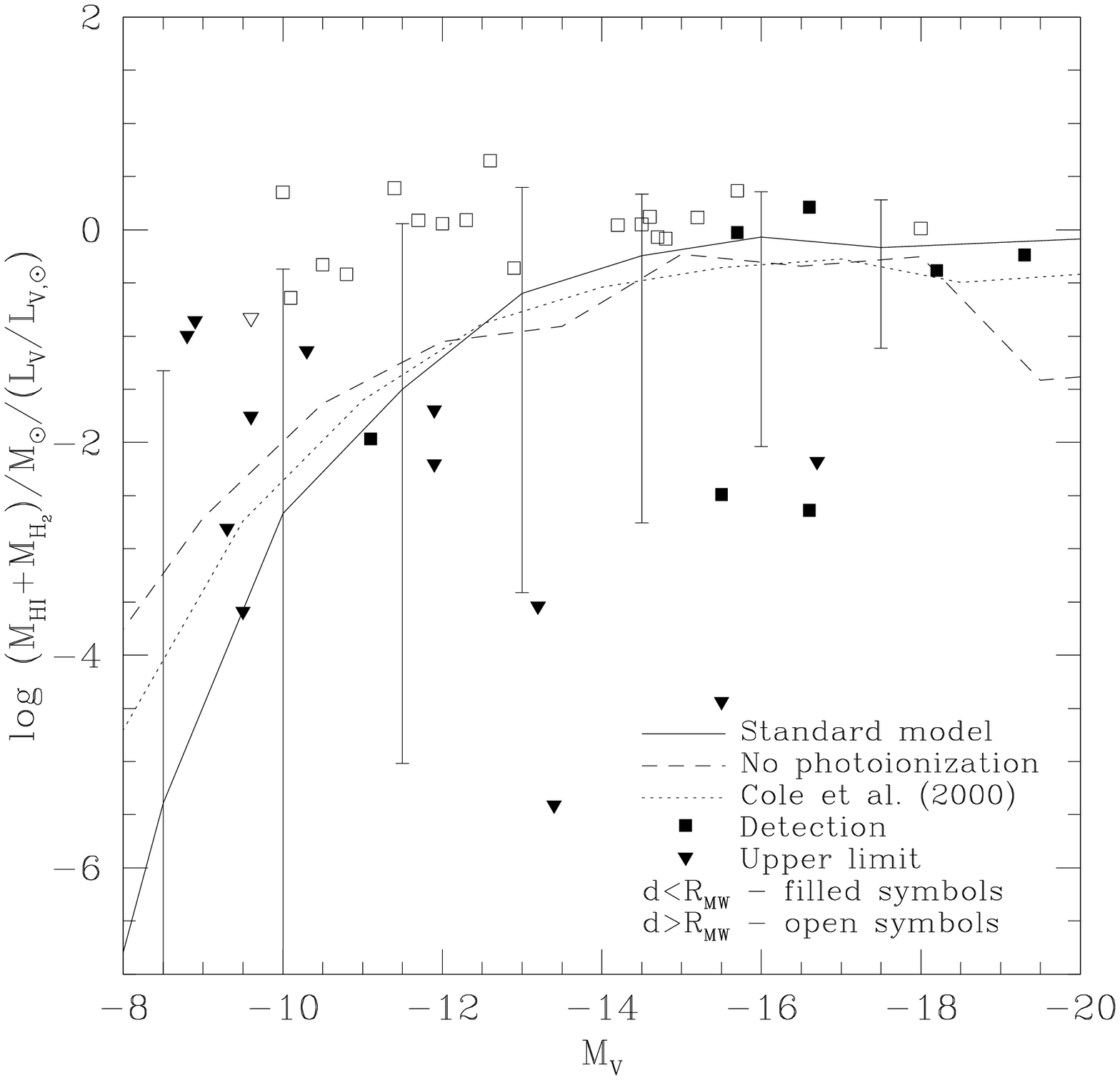,width=80mm}
\caption{The mass of hydrogen per unit V-band luminosity in satellite
galaxies as a function of V-band absolute magnitude. Data for the
Local Group satellites are taken from \protect\scite{mateo98}, except
for the LMC and SMC for which H{\sc i} masses are taken from
\protect\scite{westerlund97}, M33 for which the H{\sc i} mass comes
from \protect\scite{corbelli00} and for NGC3109 and Antlia for which
H{\sc i} masses are taken from \protect\scite{barnes01}. \HI\
detections are shown as squares, and upper limits as
triangles. Galaxies within $R_{\rm MW}$ of the Milky Way or M31 are
represented by filled symbols and more distant satellites by open
symbols. Solid, dashed and dotted lines show results from our
standard, no photoionization and \protect\scite{cole2000} models
respectively. The lines show the median model relations and the
errorbars indicate the 10\% and 90\% intervals of the distribution in
the standard model. The scatter in the other models is comparable to
this.}
\label{fig:satgas}
\end{figure}

We begin by considering the gaseous content and star formation rates 
of satellites.  The mass of hydrogen (atomic plus molecular),
normalized to the V-band luminosity, is shown, as a function of V-band
magnitude, in Fig.~\ref{fig:satgas}. A rapid decline in the gas
content towards faint magnitudes is predicted by the models. This is a
direct consequence of the strong effects of supernova feedback in
faint, low mass galaxies which efficiently eject much of their cold
gas and rapidly consume any remaining fuel (these galaxies, being
satellites, are unable to accrete fresh gas in our model). The
observational data show a band of almost constant gas-to-luminosity
ratio which is occupied exclusively by galaxies beyond $R_{\rm MW}$ of
the Milky Way or M31 at faint magnitudes. Galaxies within $R_{\rm MW}$
frequently have only upper limits to their gas mass. Our model
predictions are consistent with the observations but, since the data
often consist only of upper limits, this comparison is not
particularly conclusive.

\begin{figure}
\psfig{file=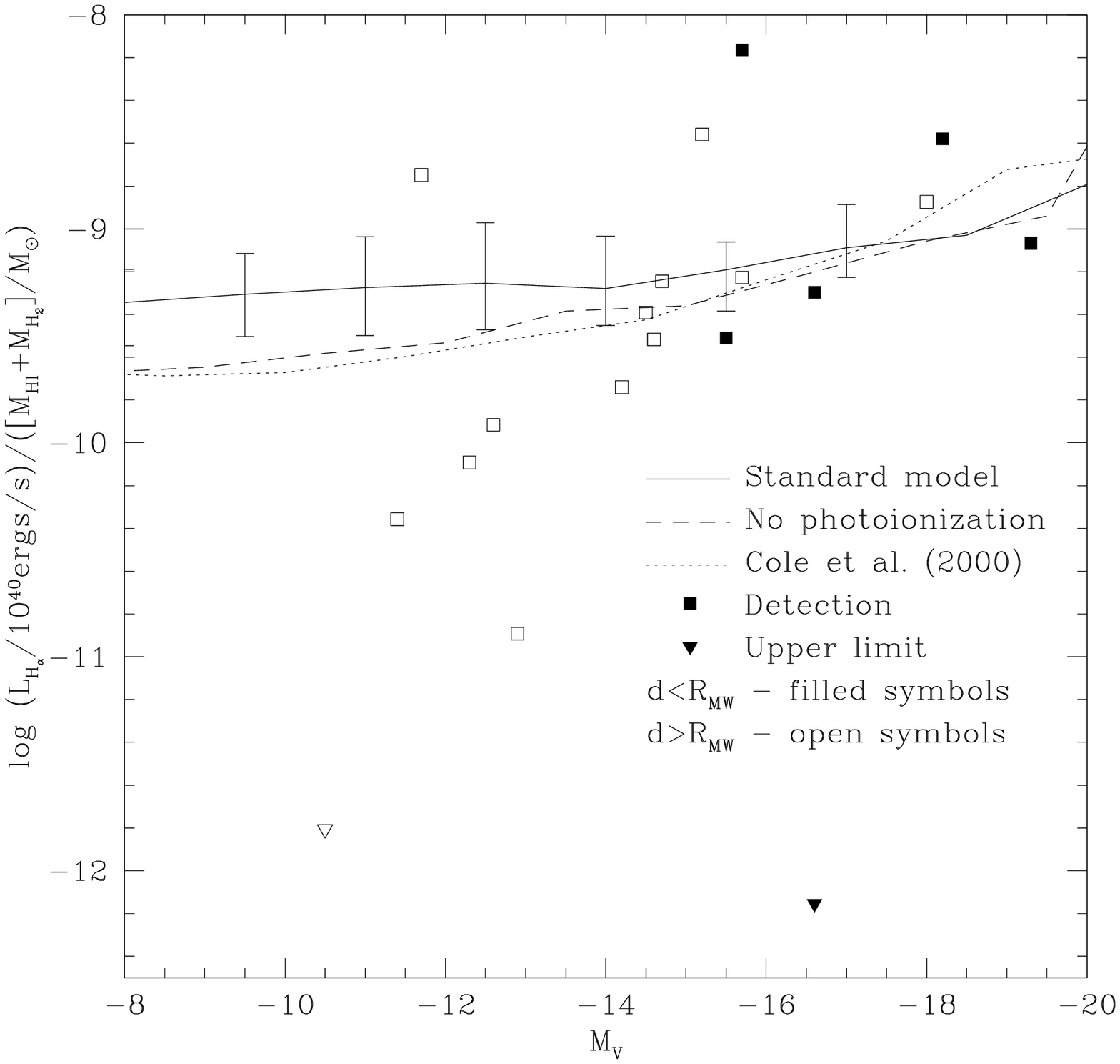,width=80mm}
\caption{The H$\alpha$ luminosity per unit mass of hydrogen for satellite
galaxies, as function of absolute V-band magnitude. Symbols show values for
Local Group satellites taken from \protect\scite{mateo98} and from
\protect\scite{westerlund97} for the Magellanic Clouds 
and \protect\scite{kennicutt98} for M33. Satellites within $R_{\rm MW}$ of
the Milky Way or M31 are shown as filled symbols, with more distant
satellites shown as open symbols. H$\alpha$ detections are denoted by
squares, and upper limits by triangles. Solid, dashed and dotted lines
indicate results from our standard, no photoionization and
\protect\scite{cole2000} models respectively. The lines indicate the
median relation and the errorbars the 10\% and 90\% intervals of the
distribution in the standard model. The scatter in the other models is
comparable to this.}
\label{fig:satHalpha}
\end{figure}

We now consider the star formation rates in satellites as estimated from
their H$\alpha$ luminosities. In Fig.~\ref{fig:satHalpha}, we plot the
H$\alpha$ luminosity per unit hydrogen mass for satellites with
measurements or limits on H$\alpha$ (all such galaxies also have
measured gas masses). With so few data points (note that many of the
galaxies in the plot lie further than $R_{\rm MW}$ from the Milky Way or
M31), it is difficult to quantify the level of agreement between model and
data. Of the six satellites within $R_{\rm MW}$ with measured H$\alpha$
luminosity, four lie close to the model prediction,
while one fainter satellite has an H$\alpha$ to hydrogen ratio an order of
magnitude above the model relation, and another has an upper limit lying
three orders of magnitude below the model expectation.

\begin{figure}
\psfig{file=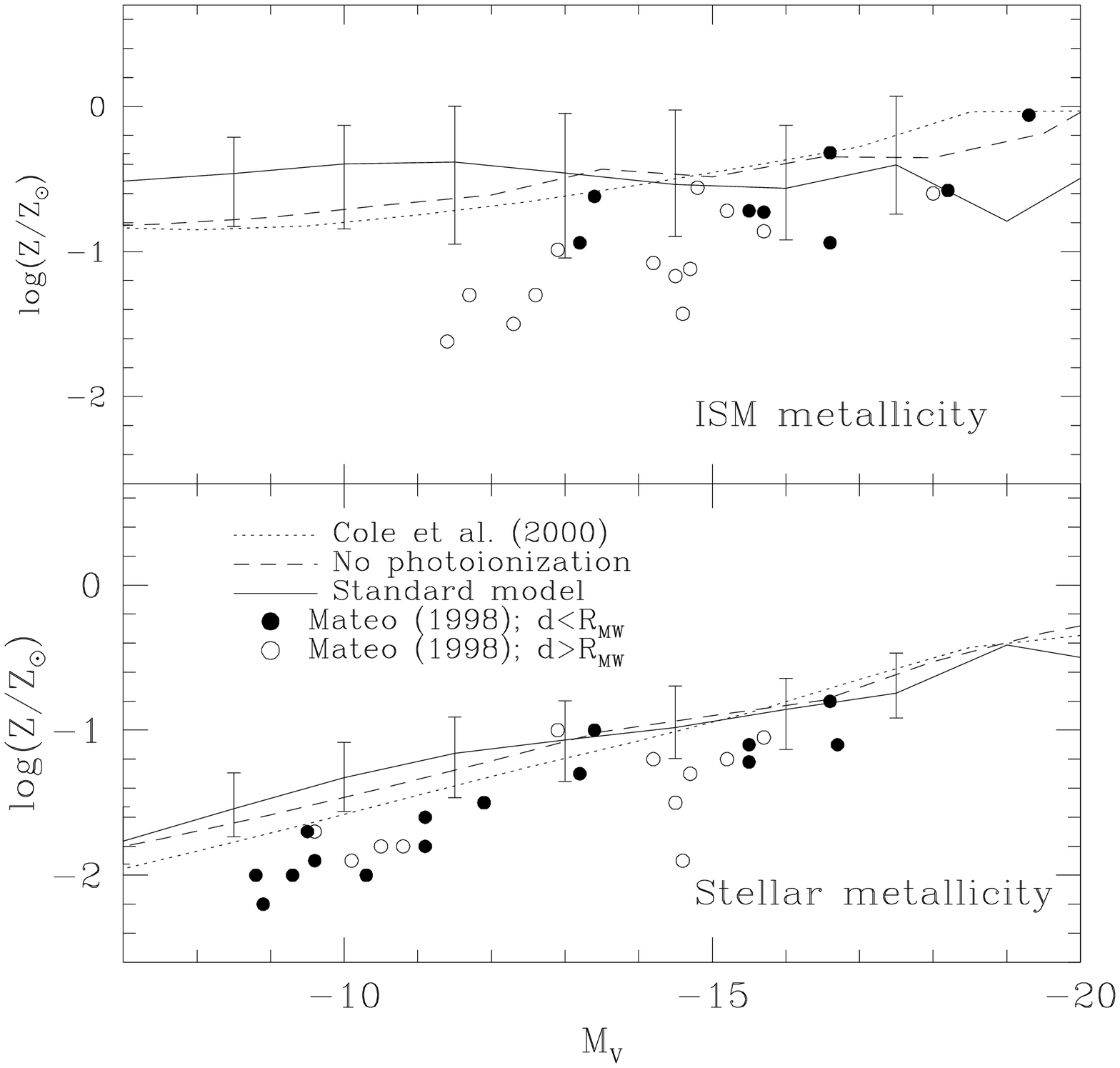,width=80mm}
\caption{The metallicity of satellite galaxies as a function of
absolute V-band magnitude.  Metallicities for Local Group satellites
are taken from \protect\scite{mateo98}, except for the Magellanic
Clouds \protect\cite{skillman89} and M33
\protect\cite{kobulnicky99}. The upper panel shows the metallicity of
gas in the ISM (relative to Solar, with $Z_\odot=0.02$), while the
lower panel shows the metallicity of the stars. Filled circles
indicate satellites within $R_{\rm MW}\approx 270$~kpc of the Milky
Way or M31, with more distant satellites shown as open circles. Solid
lines give the median relation in our standard model, with error bars
indicating the 10\% and 90\% intervals. Dashed lines correspond to our
no photoionization model (which includes tidal stripping of
satellites) and dotted lines to the \protect\scite{cole2000}
model. The scatter in these models is comparable to that in the
standard model.}
\label{fig:sat_metals}
\end{figure}

In Fig.~\ref{fig:sat_metals}, we plot the metallicity of the ISM gas (upper
panel) and of the stars (lower panel).  In both cases, the Local Group
satellites exhibit a trend of increasing metallicity with luminosity. Our
models show a trend in the same sense, although for the ISM it appears
somewhat weaker than for the data. It is interesting that the satellites
within $R_{\rm MW}$ lie much closer to the model predictions (which
are only for satellites within Milky Way-like halos) than those
outside $R_{\rm MW}$, but with the limited data available it is impossible
to say if this is a significant effect. The stellar metallicities in the
model also agree well with the data, except perhaps at the faintest
magnitudes where the model is slightly too high. While photoionization
makes little difference to the metallicities of galaxies brighter than
$M_{\rm V}\approx -15$, at fainter magnitudes the standard model predicts
slightly higher metallicities than the \scite{cole2000} model.
Photoionization causes galaxies of a given absolute magnitude to form in
higher circular velocity halos. These have deeper potential wells which
reduce the effectiveness with which supernovae feedback expels gas, thus
increasing the effective yield (see \pcite{cole2000}) and resulting in
a higher metallicity. In Paper~I, we considered the metallicity of the ISM
for galaxies in general and found that photoionization actually reduced the
gas metallicity in faint galaxies by preventing pre-processing of that gas
in smaller halos. In the case of satellites, photoionization has a much
smaller effect because, unlike central galaxies, the satellites are not
accreting gas at present, and so their metallicity is much less sensitive
to any pre-enrichment.

\begin{figure}
\psfig{file=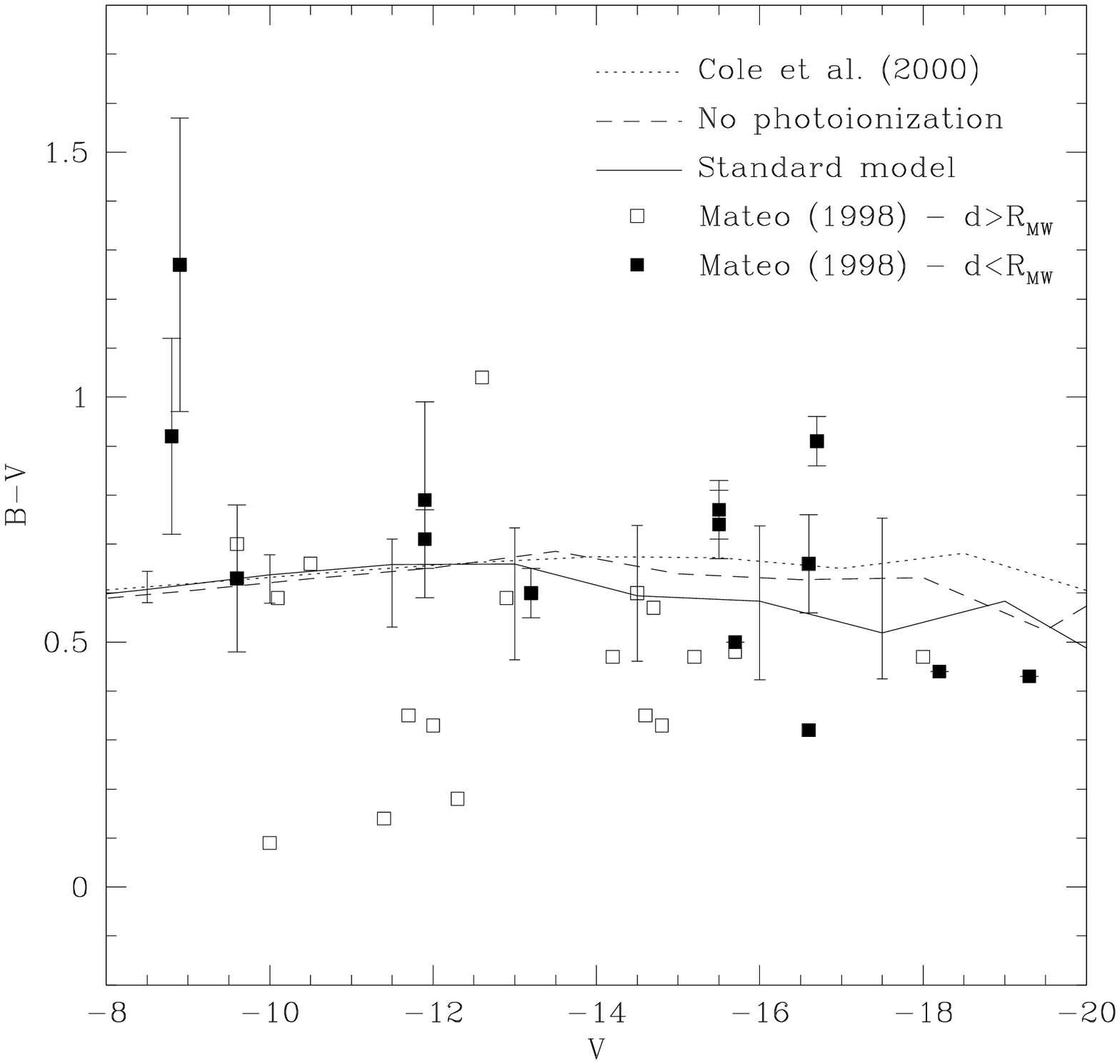,width=80mm}
\caption{The B-V colour of satellite galaxies as a function of
absolute V-band magnitude.  Colours for Local Group satellites are
taken from \protect\scite{mateo98}, except for the Magellanic Clouds
\protect\cite{skillman89} and M33 \protect\cite{kobulnicky99}. Filled
squares indicate satellites within $R_{\rm MW}\approx 270$~kpc of the
Milky Way or M31, with more distant satellites shown as open
squares. The solid line gives the median relation in our standard
model, with error bars indicating the 10\% and 90\% intervals. The
dashed line corresponds to our no photoionization model (which
includes tidal stripping of satellites) and the dotted line to the
\protect\scite{cole2000} model. The scatter in these models is
comparable to that in the standard model.}
\label{fig:sat_cols}
\end{figure}

Finally, we consider the B-V colours of satellite galaxies. These are
plotted in Fig~\ref{fig:sat_cols} for the Local Group, as a function
of satellite V-band absolute magnitude and compared with the model
predictions. The model predictions are rather insensitive to whether
or not photoionization is included.  With or without photoionization,
satellites are typically very old systems with little recent star
formation and so their colours correspond to those of an old ($\gsim
10$ Gyr), low metallicity stellar population (i.e. roughly
B-V$=$0.5--0.7 for the IMF considered here). Filled squares in the
figure indicate Local Group satellites within $R_{\rm MW}$ of the
Milky Way or M31. The colours of these galaxies are in good agreement
with the model predictions, except perhaps for two intrinsically faint
galaxies for which the observed colours are very
uncertain. Interestingly, Local Group galaxies which lie beyond
$R_{\rm MW}$ of either the Milky Way or M31 have systematically bluer
colours, suggesting that they have experienced recent star
formation. This is generically expected in our model because these
objects are still the central galaxies in their host halos and, unlike
genuine satellites, they are still able to accrete gas to fuel star
formation even at the present day. This interpreation is consistent
with the higher overall gas content measured in the most distant
galaxies as seen in Fig.~\ref{fig:satgas}

\subsection{Structure}

\begin{figure*}
\psfig{file=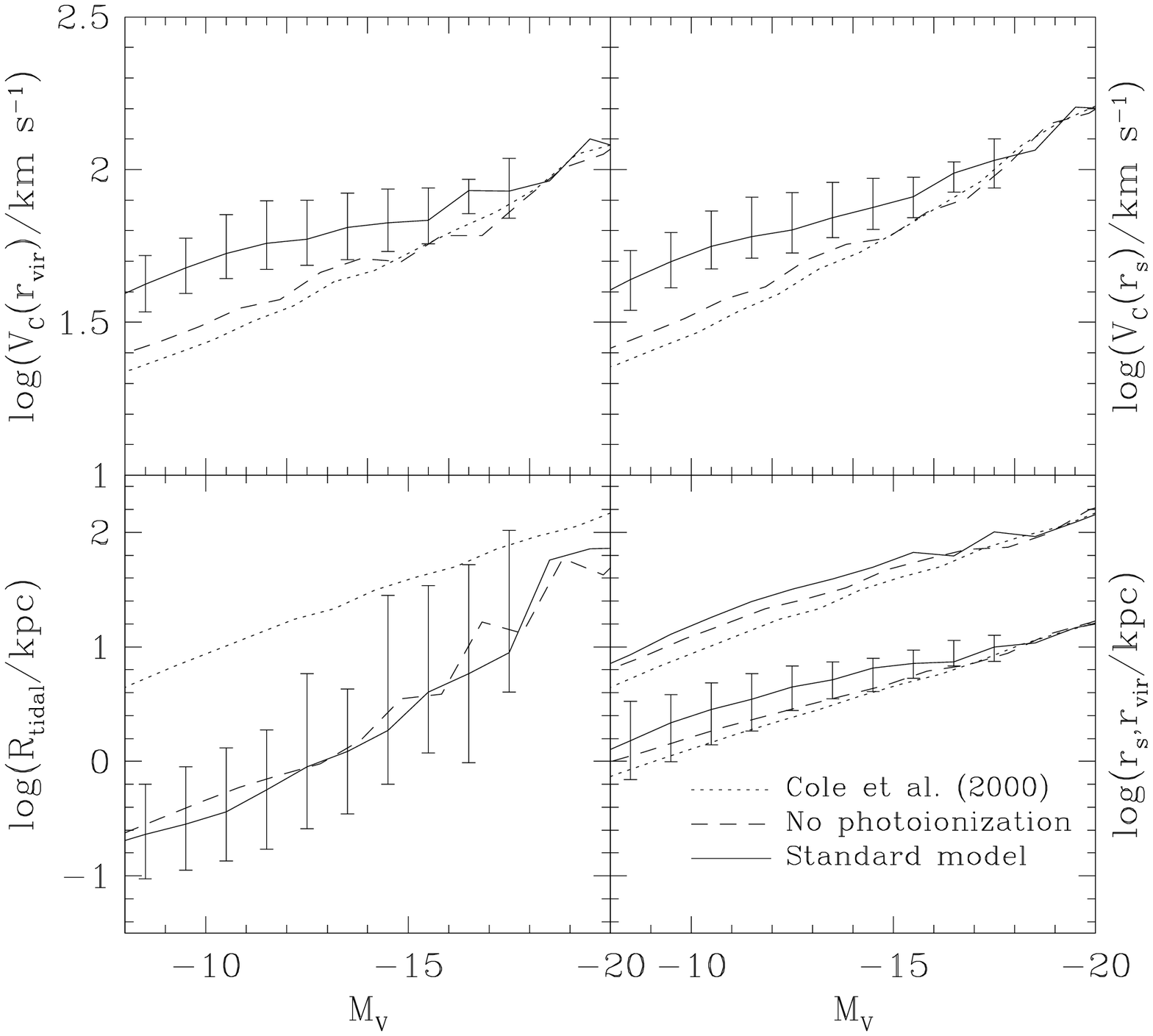,width=180mm,bbllx=0mm,bblly=90mm,bburx=200mm,bbury=270mm,clip=} 
\caption{Predicted dynamical and structural properties of satellite
galaxies and their dark matter halos, as a function of the absolute
V-band magnitude of the satellite. In all the panels, the lines show
the median model relation: solid lines for the standard model, dashed
lines for no-photoionization model (which, however, includes tidal
stripping of the satellites) and dotted lines for the 
\protect\scite{cole2000} model. As before, the errorbars indicate the
10\% and 90\% intervals of the distribution in the standard model; the
scatter in the other two models is comparable. \emph{Upper left:} The
circular velocity at the virial radius. \emph{Upper right:} The
circular velocity at the NFW scale radius, $r_{\rm s}$. This takes
into account contributions from both dark and baryonic matter in the
galaxy, including the contraction of the dark halo caused by the
condensation of the galaxy (assumed to proceed
adiabatically). \emph{Lower left:} The effective tidal radii of the
satellites. For the \protect\scite{cole2000} model, which does not
include tidal limitation, we plot instead the virial radius of the
halo for comparison. \emph{Lower right:} The NFW scale radius of the
satellite halos (lower set of lines) and the virial radius of the
halos (upper set of lines).}
\label{fig:sat_props}
\end{figure*}

Dynamical quantities describing the structure of satellite galaxies
and their dark matter halos are readily available in our model of
galaxy formation. In Fig.~\ref{fig:sat_props}, we show some of these
quantities as a function of the absolute V-band magnitude of the
satellite. We remind the reader that in our calculations, the
dynamical and structural properties of the satellite halos within the
effective tidal radius are unaffected by the mass loss beyond that
radius. In the upper panels, we plot the circular velocity at the
virial radius (left-hand panel) and at the NFW scale radius
(right-hand panel). The virial radius and corresponding circular
velocity here are the values the satellite halo last had when it was
still a separate halo, before it merged with the Milky Way halo. In
evaluating the circular velocity, we take into account contributions
from both dark and baryonic matter in the galaxy, including the
contraction of the dark halo caused by the condensation of the galaxy
(which is assumed to proceed adiabatically). For fainter satellites,
the photoionization model predicts significantly higher circular
velocities than the other models. The reason for this is one we have
encountered before: the inhibiting effect of photoionization leads to
satellites of a given luminosity forming in a more massive halo than
would be the case in the absence of photoionization. Note that such an
effect is not seen in Fig.~\ref{fig:satTF} where we plotted the
satellite galaxy ``Tully-Fisher'' relation. There, the increase in
circular velocity is offset by the reduction in the sizes of galaxies,
which causes the visible matter to sample the rotation curve at
smaller radii where the circular velocity is less.

In the lower left-hand panel of Fig.~\ref{fig:sat_props}, we plot the
effective tidal radii, as defined above and in Paper~I, of the
satellites\footnote{In our calculations, galaxies are modelled with
surface density profiles which initially extend to the virial radius
of the halo in which they formed. We can therefore define a tidal
radius for satellies even if this turns out to be much larger than the
visible extent of the galaxy. In practice, when this occurs, the tidal
radius plotted should be considered to be the tidal radius of the
satellite's halo rather than that of the satellite itself.}. For
comparison, we plot the virial radius of the halo in the
\scite{cole2000} model, which does not include tidal limitation (and
assumes, for calculations of dynamical friction, that the satellite
retains all of its halo after it merges with a large halo).  Tidal
forces strip the halos of the faintest galaxies to less than 10\% of
their initial virial radius, but this effect becomes much less
dramatic for the brighter galaxies.  (The virial radii of the halos
harbouring satellites of a given $M_{\rm V}$ are actually slightly
larger than the values given by the \scite{cole2000} model since
photoionization causes galaxies of fixed $M_{\rm V}$ to form in larger
halos).  In the lower right-hand panel, we show the NFW scale radius
of the halos (lower set of lines). For comparison, we also show the
median virial radius (i.e. the virial radius of the satellite's halo
before it merged with the Milky Way's halo; upper set of
lines). Comparison of the two shows that satellites typically formed
in halos with concentration parameters in the range $c=$5--6.

\section{Discussion}
\label{sec:discuss}

Small satellite galaxies like those that surround the Milky Way are
the descendents of some of the oldest objects in the Universe. Thus, to
understand their properties, it is necessary to take into account
processes that occured at very early times, like the reionization of
the Universe, which are often neglected when studying the properties
of larger galaxies. Furthermore, since these satellites orbit in the
halo of the parent galaxy, it is also important to take into account
dynamical processes such as tidal effects and dynamical friction. We
have used an extension of the \scite{cole2000} semi-analytic model of
galaxy formation which includes all these processes to study the
expected abundance and properties of satellite galaxies in the Local
Group. Our work improves upon earlier studies by \scite{kwg93} and
\scite{bullock00} because it self-consistently calculates the physics
of reionization and galaxy formation and includes a treatment of the
main dynamical effects experienced by satellites orbiting in a dark
matter halo.

We generated samples of halos containing galaxies similar to the Milky
Way.  Photoionization, which occurs at $z\approx 8$, inhibits the
formation of small galaxies and so the satellites that survive to the
present tend to be those that formed while the universe was still
neutral. This has the consequence of greatly suppressing the number of
satellites of a given luminosity relative to the number that would be
expected if photoionization were neglected. Furthermore, at a given
satellite luminosity, we find that it is those satellites with the
lowest circular velocity that are preferentially depleted by the
effects of photoionization. The measured distribution function of
satellite circular velocity depends on this differential destruction
but also on the internal structure of the satellite which determines
the shape of its rotation curve. We take the `measured' circular
velocity of a satellite to be the value at the half-light radius. We
find that for the fainter satellites ($M_{\rm V}\simeq -10$), the
measured circular velocity is about half the value at the virial
radius because the half-light radius is within the rising part of the
rotation curve, whereas for the brighter satellites ($M_{\rm V}\simeq
-15$), the measured circular velocity is very similar to the value at
the virial radius. These effects work together to produce both a
luminosity function and a circular velocity function which are in good
agreement with the available data. This result is remarkable because
we have not had to adjust a single parameter value in the
\scite{cole2000} model nor fine-tune our treatment of photoionization.

Our model also predicts the sizes and metallicities of satellites and
these also turn out to be in good agreement with the limited amount of
data currently available.  The model predicts that photoionization has
negligible effects in galaxies with circular velocity above about
60km\,s$^{-1}$, in agreement with earlier analytical
\cite{efstath92,thoul96} and numerical treatments
\cite{quinn96,navarro97}.

Our model can, in principle, be tested on the basis of the predictions
it makes for as yet unobserved properties of the satellite population
in the Local Group. In particular, our model suggests that there
should be a large population of faint satellites around the Milky Way
awaiting discovery.  A near complete census would require deep imaging
(at least to 26 V-band magnitudes per square arcsecond). Alternatively,
satellites may be recognized as an excess of stars against the
background. We have presented detailed predictions for the expected
number of stars and their surface density as a function of the
luminosity of the satellite to which they belong. These calculations
may be useful in designing observational strategies aimed at
discovering new satellites.  A further possibility is to try and
detect these satellites by searching for their \HI\ content
\cite{putman}. Our model predicts that $M_{\rm V}=-10$ satellites
should typically contain $10^5M_\odot$ of \HI, with a rapid decline in
\HI\ mass at  fainter magnitudes. Such observations are difficult, but 
perhaps not impossible. 

We conclude that speculative mechanisms such as non-standard inflation
\cite {kamion99} or new components of the Universe such as warm,
self-interacting or annihilating dark matter
\cite{hogan99,spergel99,yoshida00,craigdavis01} are not required to
explain the observed abundance of Local Group satellites. Instead, the
low abundance of satellites is a natural consequence of galaxy
formation in a CDM universe when the physical effects of
photoionization and tidal interactions, two processes which are known
to occur, are taken into account.  Continuing improvement in the
observational data for Local Group galaxies, particularly better
measurements of their structure and dynamical state and an assessment
of the completeness of existing samples, together with searches for
new satellites, will provide a strong test of current models of galaxy
formation.

\section*{Acknowledgments}

CSF acknowledges a Leverhulme Research Fellowship. CGL acknowledges
support at SISSA from COFIN funds from MURST and funds from ASI. CMB
acknowledges a Royal Society University Research Fellowship. SMC
acknowledges a PPARC Advanced Fellowship. We also thank Ben Moore for
drawing our attention to the Milky Way satellite problem.

\end{document}